\documentclass[12pt
	       ,final
               ]{article}
\usepackage{amsmath,amssymb}
\usepackage[final]{epsfig}
\usepackage{hyperref}

\newcommand{\draftnote}[1]{}

\let\origcaption=\caption
\renewcommand{\caption}[1]{\origcaption{\small #1}}

\newcommand{\Tr}{\mathop\mathrm{Tr}\nolimits}
\newcommand{\Real}{\mathop\mathrm{Re}\nolimits}
\newcommand{\Imag}{\mathop\mathrm{Im}\nolimits}
\newcommand{\Dm}{D_\mu}
\newcommand{\Dmu}{D^\mu}
\newcommand{\dm}{\partial_\mu}
\newcommand{\dn}{\partial_\nu}
\newcommand{\cB}{\mathcal{B}}

\newcommand{\bra}[1]{\langle #1 |}
\newcommand{\ket}[1]{| #1 \rangle}
\newcommand{\bm}[1]{\boldsymbol{#1}}
\newcommand{\bx}{{\bm{x}}}
\newcommand{\bk}{{\bm{k}}}
\newcommand{\bkind}{^{\vphantom{*}}_\bk}

\newcommand{\fk}{f^{\vphantom{*}}_\bk}
\newcommand{\fmk}{f^{\vphantom{*}}_{-\bk}}
\newcommand{\gsk}{g^{*}_\bk}
\newcommand{\half}{\frac{1}{2}}
\newcommand{\e}{\mathrm{e}}
\newcommand{\const}{\mathrm{const}}

\newcommand{\vpar}[2]{\frac{\delta #1}{\delta #2}}

\providecommand{\href}[2]{#2}

\begin{document}

\title{Instanton--Like Processes in Particle Collisions:  a Numerical
  Study of the $SU(2)$-Higgs Theory below the Sphaleron Energy.}
\author{F.~Bezrukov$^a$%
  \thanks{\protect\href{mailto:fedor@ms2.inr.ac.ru}%
                      {\texttt{fedor@ms2.inr.ac.ru}}},\and
        C.~Rebbi$^b$%
  \thanks{\protect\href{mailto:rebbi@bu.edu}%
                      {\texttt{rebbi@bu.edu}}},\and
        V.~Rubakov$^a$%
  \thanks{\protect\href{mailto:rubakov@ms2.inr.ac.ru}%
                      {\texttt{rubakov@ms2.inr.ac.ru}}},\and
        P.~Tinyakov$^{c,a}$%
  \thanks{\protect\href{mailto:Peter.Tinyakov@cern.ch}%
                      {\texttt{Peter.Tinyakov@cern.ch}}}%
  \and
  \small\itshape (a) Institute for Nuclear Research,\\
    \small\itshape 60th October Anniversary pr. 7a, Moscow 117312, Russia\\
  \small\itshape (b) Department of Physics---Boston University\\
    \small\itshape 590 Commonwealth Avenue, Boston MA 02215, USA\\
  \small\itshape (c) Institute of Theoretical Physics, University of
          Lausanne,\\
    \small\itshape CH-1015 Lausanne, Switzerland
}
\date{}

\maketitle

\begin{abstract}
We use semiclassical methods to calculate the probability of inducing
a change of topology via a high-energy collision in the $SU(2)$-Higgs
theory.  This probability is determined by a complex solution to a
classical boundary value problem on a contour in the complex time
plane.  In the case of small particle number it is the probability of
instanton-like processes in particle collisions.  We obtain
numerically configurations with the correct topological features and
expected properties in the complex time plane.  Our work demonstrates
the feasibility of the numerical approach to the calculation of
instanton-like processes in gauge theories.  We present our
preliminary results for the suppression factor of topology changing
processes, which cover a wide range of incoming particle numbers and
energies below the sphaleron energy.
\end{abstract}

\section{Introduction and Summary}
\label{sec:intro}

Non-perturbative effects occur in many processes studied by quantum field
theory.  Well known examples are the  decay of the false vacuum in scalar 
models and instanton-like transitions in gauge theories.  The latter are
accompanied by non-\hspace{0pt}conservation of fermion quantum numbers
\cite{'tHooft:1976fv}, such as baryon number, and thus are of
importance for particle physics and cosmology.

In weakly coupled theories, these processes are described \emph{at low
energies} by classical Euclidean solutions interpolating between
initial and final states separated by an energy barrier.  In the examples
mentioned above such solutions are known as bounce~\cite{Coleman:1977py} 
and instanton~\cite{Belavin:1975fg}.  The Euclidean action of the solution 
determines the exponential part of the transition rate.  It is inversely
proportional to a small coupling constant present in the model and
thus, in general, the rate is highly suppressed.

At energies comparable to the height of the barrier, the probabilities
of the transitions between topologically distinct vacua may become
unsuppressed.  This takes place, for example, at finite temperature
\cite{Kuzmin:1985mm,Arnold:1987mh,Arnold:1988zg}, finite fermion
density \cite{Rubakov:1985am,Matveev:1987gq,Deryagin:1986kx}, or in
the presence of heavy fermions in the initial state
\cite{Rubakov:1985ix,Ambjorn:1985bb,Rubakov:1985it}.  In high energy
particle collisions the situation is not quite the same.

As was first noted in \cite{Ringwald:1990ee,Espinosa:1990qn}, at
relatively low energies the corrections to the tunneling rate can be
calculated by perturbative expansion in the background of the
instanton (bounce).  Further studies showed that the actual expansion
parameter is $\varepsilon=E/E_\mathrm{sph}$
\cite{McLerran:1990ab,Khlebnikov:1991ue,Yaffe:1990iy,Arnold:1991cx}
and the total cross section of induced tunneling has an
exponential form
\begin{displaymath}
  \sigma_{tot} (E) \sim \exp \left\{
    -\frac{16\pi^2}{g^2} F_{HG}(E/E_\mathrm{sph}) \right\} ,
\end{displaymath}
where $g$ is the small coupling constant and the function
$F_{HG}(\varepsilon)$ is a series in powers of
$\varepsilon\equiv E/E_\mathrm{sph}$
(for a review see~\cite{Tinyakov:1993dr}).

While the perturbation theory in $\varepsilon$ is limited to small
$\varepsilon$, the general form of the total cross section implies
that there might exist a semiclassical-type procedure which would
allow, at least in principle, to calculate $F_{HG}(\varepsilon)$ at
$\varepsilon\gtrsim 1$.  Since the initial state of two highly
energetic particles is not semiclassical, the standard semiclassical
procedure does not apply and a proper generalization is needed, which
was proposed in
refs.~\cite{Rubakov:1992fb,Tinyakov:1992fn,Rubakov:1992ec}.  The
corresponding formalism reduces the calculation of the exponential
suppression factor to a certain classical boundary value problem,
whose analytical solution is not usually possible.

The semiclassical approach proposed in
refs.~\cite{Rubakov:1992fb,Tinyakov:1992fn,Rubakov:1992ec} is based on
the conjecture that, with exponential accuracy, the two-particle
initial state can be substituted by a multiparticle one provided that
the number of particles is not parametrically large.  The few-particle
initial state, in turn, can be considered as a limiting case of truly
multiparticle one with the number of particles $N=\nu/g^2$ when the
parameter $\nu$ is sent to zero. For the multiparticle initial state
the transition rate is explicitly semiclassical and has the form
\[
  \sigma(E,N) \sim \exp \left\{
    -\frac{16\pi^2}{g^2} F(\varepsilon,\nu) \right\} .
\]
According to the above conjecture, the function $F_{HG}(\varepsilon)$,
corresponding to the two-particle incoming state, is reproduced in the
limit $\nu\to 0$,
\[
  \lim\limits_{\nu\to 0}F(\varepsilon,\nu) = F_{HG}(\varepsilon).
\]
Therefore, although indirectly, the function $F_{HG}(\varepsilon)$ is
also calculable semiclassically.  Although not proven rigorously, this
conjecture was checked explicitly in field theory models in several
orders of perturbation theory in $\varepsilon$~\cite{Mueller:1993sc}
and also in a quantum-mechanical model for all
$\varepsilon$~\cite{Bonini:1999kj}.

Until now the only analyses of induced tunneling have been performed in a
quantum mechanical model \cite{Bonini:1999kj} and in a scalar field
model of false vacuum decay \cite{Kuznetsov:1997az}.  A particularly
interesting case to study is, however, the Electroweak Theory where
different topological sectors are separated by a potential barrier
of the height $E_\mathrm{sph}\sim10\mathrm{TeV}$
\cite{Manton:1983nd,Klinkhamer:1984di}.  Whether the exponential
suppression disappears in this theory at sufficiently high energy is
still an open question.  In this paper we consider an $SU(2)$ gauge
theory with Higgs doublet, which corresponds to the Electroweak sector
of the Standard Model with $\theta_W=0$.

Classically allowed over-barrier sphaleron transitions in this model
were studied in ref.~\cite{Rebbi:1996zx}.  All solutions found in
ref.~\cite{Rebbi:1996zx} are configurations with large number of
particles in the initial state and thus they do not correspond to
few-particle collisions.

In the present work we adapt the prescription of
\cite{Rubakov:1992fb,Tinyakov:1992fn,Rubakov:1992ec} to
theories with gauge degrees of freedom.  The prescription requires 
the solution of field
equations on a contour in complex time plane (see
figure~\ref{fig:time_contour}) with boundary conditions of a special
form.  Since the solutions cannot be found analytically, one has to
invoke numerical techniques.  Here we present solutions to this
problem in a limited region of parameters.  Our solutions possess
all expected features, including correct topological structure and
singularity structure in complex time plane.  Thus, our results
appear to validate the use of numerical methods
for the study of instanton-like transitions in gauge
theories at high energies.

In this work we explore the region of parameters with $E<E_\mathrm{sph}$ and
$0.4\lesssim\nu<1$.  We calculate the exponent of the
suppression factors for a wide range of values of $E$ and
$\nu$ within the above region.  Not surprisingly, since
we work at energies below the sphaleron energy, we find that
topology changing processes remain exponentially suppressed
throughout the region we studied.  More relevant may be the
fact that the trend of our results appears to indicate that
topology changing processes with low initial particle number
may remain suppressed well above $E_\mathrm{sph}$.

The paper is organized as follows.  In section \ref{sec:bvp} we briefly
review the
boundary value problem for the semiclassical calculation of
$\sigma(E,N)$. In section \ref{sec:model} 
we present the $SU(2)$ model used in our calculations.  In section
\ref{sec:num_results} we describe our numerical results and
section \ref{sec:conclusions} contains a few concluding
remarks.

\section{Semiclassical approach to induced tunneling probability}
\label{sec:bvp}

The inclusive multiparticle probability of a transition from a state with
fixed energy $E$ and number of particles $N$ about one vacuum to
\emph{any} state about another vacuum can be written in the form:
\begin{displaymath}
  \sigma(E,N) =
    \sum_{i,f} |\bra{f} \hat{S} \hat{P}_E \hat{P}_N \ket{i}|^2 \;,
\end{displaymath}
where $\hat{S}$ is the $S$-matrix, $\hat{P}_{E,N}$ are projectors onto
subspaces of fixed energy $E$ and fixed number of particles $N$, and
the states $\ket{i}$ and $\ket{f}$ are perturbative excitations about
topologically distinct vacua.
The method of semiclassical calculation of this
probability $\sigma(E,N)$ was formulated in refs.~%
\cite{Rubakov:1992fb,Tinyakov:1992fn,Rubakov:1992ec,Kuznetsov:1997az}.
Here we only review the prescription.

\begin{figure}
  \centerline{%
    \begin{picture}(0,0)%
      \psfig{file=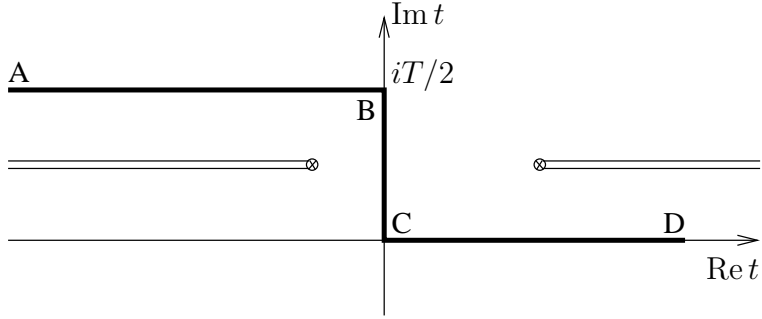}%
    \end{picture}%
    \setlength{\unitlength}{4144sp}%
    \begin{picture}(4545,1923)(418,-1423)
      \put(2746,-61){\makebox(0,0)[lb]{$iT/2$}}
      \put(2746,344){\makebox(0,0)[lb]{$\Imag t$}}
      \put(4951,-1186){\makebox(0,0)[rb]{$\Real t$}}
    \end{picture}%
  }
  \caption{The contour in complex time plane used in the formulation
    of the boundary value problem \eqref{final_BVP}.  Crossed circles
    represent singularities of the field at $r=0$, for larger $r$ the
    singularities generally move to larger times.}
  \label{fig:time_contour}
\end{figure}

For small coupling constant $g$, the semiclassical approximation is
applicable.  It boils down to the classical boundary value problem
specified on the contour in complex time plane shown in
figure~\ref{fig:time_contour}.  In the initial part of the contour
(AB) at sufficiently large negative times the fields have to be in the
linear regime.  Let us denote the fields collectively as
$\varphi(\bx,t)$.  The frequency components at distant past (part A of
the contour), $\fk$ and $g\bkind$, are defined as follows,
\[
  \varphi({\bm x},t)\big|_{T_i} =
   \int \frac{d\bk}{ (2\pi)^{3/2}{\sqrt{2\omega_{\bk}}} }
   {\left( \fk \e^{-i\omega_\bk(t-iT/2)+i{\bm{kx}}}
    + g_{\bk}^* \e^{i\omega_\bk(t-iT/2)-i{\bm{kx}}}  \right)}    
\]
The tunneling probability reads (after proper rescaling of the fields)
\begin{gather}
  \label{PIsigmaE,N}
  \sigma(E,N) \sim
  \exp \left\{ - \frac{16\pi^2}{g^2} F(\varepsilon,\nu) \right\} \\
  \label{Fexp}
  - \frac{16\pi^2}{g^2} F(\varepsilon,\nu) =
  N\Theta + ET - 2\Imag [S_{ABCD}(\varphi)]+\Real\cB_i \;,
\end{gather}
where
\[
  \cB_i=\frac{1}{2}\int\! d\bk(\fk \fmk\e^{-2i\omega_k(T_i-iT/2)}
        -g_{\bk}^* g_{-\bk}^*\e^{2i\omega_k(T_i-iT/2)})
\]
(it is easy to check that the expression \eqref{Fexp} for
$F(\varepsilon,\nu)$ is independent of $T_i$ if the system is in
linear regime at initial time).  Here the field $\varphi$ interpolates
between neighborhoods of topologically distinct vacua and satisfies
the field equation:
\begin{subequations}\label{final_BVP}
\begin{equation}
  \label{BVP_eq}
  \vpar{S}{\varphi} = 0
\end{equation}
At initial time the frequency components of the solution should
satisfy the following equations (``$\Theta$ boundary condition'')
\begin{equation}\label{BVP_bc2}
  \fk = \e^{-\Theta} g_{\bk} \;.
\end{equation}
On the final part of the contour (CD), the field is real, so that
\begin{equation}\label{BVP_real}
  \Imag \dot \varphi(\bm{x},0) = \Imag \varphi(\bm{x},0) = 0 \;.
\end{equation}
\end{subequations}
Equations \eqref{BVP_eq}--\eqref{BVP_real} specify the boundary value
problem corresponding to the induced topological transition.

The quantities entering equation \eqref{Fexp} are defined as follows.
$S_{ABCD}$ is the action for the solution of equations
\eqref{final_BVP}, energy and number of incoming particles are
\begin{equation}
  \label{BVP_EN}
  E = \int d\bk\, \omega\bkind \fk\gsk \;,\qquad
  N = \int d\bk\, \fk\gsk \;.
\end{equation}
These equations indirectly fix the values of the auxiliary variables
$T$ and $\Theta$ for given energy and number of particles.
Alternatively, one can fix $T$ and $\Theta$, solve the boundary value
problem \eqref{final_BVP} and obtain the corresponding values of $E$
and $N$ using \eqref{BVP_EN}.  This is especially convenient in
numerical calculations.

The interpretation of solutions to the boundary value problem
(\ref{final_BVP}) is as follows.  On the part CD of the contour the
saddle-point field is real; it describes the evolution of the system
after tunneling.  On the contrary, it follows from boundary conditions
(\ref{BVP_bc2}) that in the initial asymptotic region, the
saddle-point field is complex provided that $\Theta\neq 0$.  Thus, the
initial state which maximizes the probability (\ref{PIsigmaE,N}) is
not described by a real classical field, i.e.\ the classical field
must be analytically continued to complex values and this stage of the
evolution is essentially quantum even at $N\sim 1/g^2$.

The picture described implies that there exist singular points of the
solution in the complex time plane, as shown in
figure~\ref{fig:time_contour}.  To understand this, one notices that
on the negative part of the real time axis, the solution (at least for
energies below the sphaleron energy) ``bounces back'' to the same
vacuum, as in the CD part.  On the other hand, the solution in the AB
part of the contour and its analytic continuation to the real axis is
close to a different vacuum.  This may happen only if a branch cut
exists between the real axis and AB part of the contour.  Similar
arguments require a singularity to exist to the right of the BC part
of the contour too.

The case $\Theta=0$ is exceptional. In this case, the boundary
condition (\ref{BVP_bc2}) reduces to the reality condition imposed at
$\Imag t =T/2$. The solution to the resulting boundary value problem
is a periodic instanton of ref.\cite{Khlebnikov:1991th}.  The periodic
instanton is a real periodic solution to the Euclidean field equations
with period $T$ and two turning points at $t=0$ and $t=iT/2$. Being
analytically continued to the Minkowskian domain through the turning
points, periodic instanton is real at the lines $\Imag t =0$ and
$\Imag t = T/2$ and therefore satisfies the boundary value problem
(\ref{final_BVP}) with $\Theta=0$. Periodic instanton solutions
have been studied with a computational approach similar
to the one used in this paper in ref.\cite{Bonini:1999fc}.

\section{The model}\label{sec:model}

In this paper we study the four-dimensional model which captures all
the important features of the Standard Model---an $SU(2)$ gauge theory
with the Higgs doublet.  This model corresponds to the bosonic sector
of the Standard Model with $\theta_W=0$.  Also, according to
ref.~\cite{Espinosa:1991fc} we ignore the back reaction of fermions on
the gauge and Higgs fields dynamics.  The action of the model is
\begin{equation}\label{SU2action}
  S = \int d^4x \left\{\vphantom{\frac{1}{2}}
      -\frac{1}{2}\Tr F_{\mu\nu}F^{\mu\nu}
      +(\Dm\Phi)^\dagger\Dmu\Phi
      -\lambda(\Phi^\dagger\Phi-1)^2
      \right\} \;,
\end{equation}
where
\begin{align*}
  F_{\mu\nu} &= \dm A_\nu-\dn A_\mu-i[A_\mu,A_\nu] \\
  \Dm\Phi    &= (\dm-iA_\mu)\Phi
\end{align*}
with $A_\mu=A_\mu^a\sigma^a/2$.  Here we have eliminated some
inessential constants by an appropriate choice of units. We have also
set the gauge coupling constant $g=1$ by proper rescaling of the
fields and action, but it can be easily restored in the final result
(see \eqref{Fexp}).  The Higgs self-coupling $\lambda$ was set equal
to $\lambda=0.125$ in all calculations, which corresponds to
$m_H=m_W$.

This theory is still too complicated for numerical study because of
large number of variables.  To make the problem computationally manageable
we consider only configurations spherically symmetric in space
\cite{Ratra:1988dp}, which reduces the system to an effective
2-dimensional theory.  It still possesses many features of the full
4-dimensional model, such as a similar topological structure.
Moreover, for large times the energy disperses along the
radial direction as it would do in the full 4-dimensional theory.
This guarantees that the system linearizes
with time making it possible to impose boundary conditions
\eqref{BVP_bc2} in the asymptotic region.

The spherical \emph{Ansatz} is given by expressing the fields in terms
of six real functions $a_0$, $a_1$, $\alpha$, $\beta$, $\mu$ and
$\nu$:
\begin{subequations}\label{ansatz}
\begin{align}
  A_0(\bm{x},t) &= \half a_0(r,t) \bm{\sigma\cdot n} \\
  A_i(\bm{x},t) &= \half\left[a_1(r,t)\bm{\sigma\cdot n}n_i
                   +\frac{\alpha(r,t)}{r}(\sigma_i
                                          -\bm{\sigma\cdot n}n_i)
                   +\frac{1+\beta(r,t)}{r}\epsilon_{ijk}n_j\sigma_k
                   \right] \\
  \Phi(\bm{x},t) &= [\mu(r,t)+i\nu(r,t)\bm{\sigma\cdot n}]\xi \;,
\end{align}
\end{subequations}
where $\bm{n}$ is the unit three-vector in the radial direction and
$\xi$ is an arbitrary constant two-component complex unit vector.  The
action \eqref{SU2action} expressed in terms of the new fields becomes
\begin{displaymath}
\begin{split}
  S = 4\pi\int dt \int_0^\infty dr
      \left[\vphantom{\half}\right. &
        \frac{1}{4}r^2f_{\mu\nu}f_{\mu\nu}
        + (\bar D_\mu \bar \chi)D_\mu \chi
        + r^2 (\bar D_\mu \bar\phi)D_\mu\phi \\
      & -\frac{1}{2 r^2}\left( ~\bar\chi\chi-1\right)^2
        -\frac{1}{2}(\bar\chi\chi + 1)\bar\phi\phi \\
      & \left.\vphantom{\half}
        -\frac{i}{2} \bar\chi \phi^2+\frac{i}{2} \chi \bar\phi^2
        -{\lambda}  r^2 (\bar\phi\phi- 1)^2 \right]
\end{split}
\end{displaymath}
where the indices $\mu$, $\nu$ run from 0 to 1 and
\begin{subequations}\label{defns}
\begin{align}
  && f_{\mu\nu}=\partial_\mu a_\nu - \partial_\nu a_\mu & \\
       \chi &=\alpha+i\beta&
   \bar\chi &= \alpha - i\beta\\
       \phi &= \mu + i \nu&
   \bar\phi &= \mu - i \nu\\
  D_\mu\chi &= (\partial_\mu-i a_\mu)\chi&
  \bar D_\mu\bar\chi &= (\partial_\mu + i a_\mu)\bar\chi\\
  D_\mu \phi &= (\partial_\mu- \frac{i}{2} a_\mu)\phi&
  \bar D_\mu \bar\phi &= (\partial_\mu + \frac{i}{2} a_\mu)\bar\phi\;.
\end{align}
\end{subequations}
Note that the overbar on $\phi$, $\chi$ and $D_{\mu}$ denotes changing
$i \to -i$ in the definitions (\ref{defns}) above, which is the same
as complex conjugation \emph{only} if the six fields $a_{\mu}$,
$\alpha$, $\beta$, $\mu$ and $\nu$ are real.  In the boundary value
problem \eqref{final_BVP} these fields become complex and overbar no
longer corresponds to normal complex conjugation.

\paragraph{Vacuum structure.}
The spherical \emph{Ansatz} (\ref{ansatz}) has a residual $U(1)$
gauge invariance
\begin{align*}
  a_\mu &\to a_\mu + \partial_\mu \Omega \\
  \chi  &\to \e^{i \Omega} \chi \\
  \phi  &\to \e^{i \Omega/2} \phi  \;,
\end{align*}
with gauge function $\Omega(r,t)$.  The complex ``scalar'' fields
$\chi$ and $\phi$ have $U(1)$ charges $1$ and $1/2$ respectively,
$a_{\mu}$ is the $U(1)$ gauge field, $f_{\mu\nu}$ is the field
strength, and $D_{\mu}$ in \eqref{defns} is the covariant derivative.
The residual $U(1)$ gauge invariance must be fixed when solving the
equations numerically.  In our work we chose the temporal gauge $a_0 =
0$ and impose Gauss' law (the equation corresponding to variation over
$a_0$) at the initial moment of time.

The trivial space-independent vacuum of the model has the form
\begin{displaymath}
  \chi_{\mathrm{vac}}=-i\;,\quad
  \phi_{\mathrm{vac}}=\pm 1\;,\quad
  a_{1\,\mathrm{vac}}=0\;.
\end{displaymath}
Other vacua can be obtained from the trivial one by the gauge
transformations:
\begin{align*}
  a_{\mu\,\mathrm{vac}} &= \dm\Omega \\
  \chi_\mathrm{vac}     &= -i\e^{i\Omega} \\
  \phi_\mathrm{vac}  &= \pm\e^{i\Omega/2} \;.
\end{align*}
$\Omega$ should be zero at origin.  Vacua with different winding
numbers correspond to $\Omega\to2n\pi$ as $r\to\infty$.  For such
values of $\Omega$, the fields of the original four-dimensional model
are constant at spatial infinity, which is the standard choice.  It
allows for a standard description of the topological properties of
vacua---since the sphere $S^2$ at spatial infinity is mapped to one
point in field space, one can compactify the space to $S^3$ and
consider mappings $S^3\to SU(2)$ (the latter correspond to pure gauge
field configurations).

One can also make any other choice of fields at spatial infinity (as
long as the fields are pure gauge and constant in time there).  In our
case it is convenient to set $\Omega\to(2n-1)\pi$ at $r\to\infty$.
This is equivalent to the requirement that the fields satisfy the
following boundary conditions at the ``boundaries'' of space:
\begin{align*}
  \chi|_{r\to0}                   &\to -i &    \chi|_{r\to\infty} &\to
  i \nonumber\\
  \label{space_bc}
  \Real\partial_r\phi|_{r\to0} &\to 0  & \phi|_{r\to\infty} &\to i \\
  \Imag\phi|_{r\to0}           &\to 0 \;. && \nonumber
\end{align*}
In the original 4-dimensional theory this means that the sphere $S^2$
at spatial infinity is mapped onto the equatorial sphere of $SU(2)$
parameterizing the gauge vacua.

\begin{figure}
  \centerline{%
    \begin{picture}(0,0)%
      \psfig{file=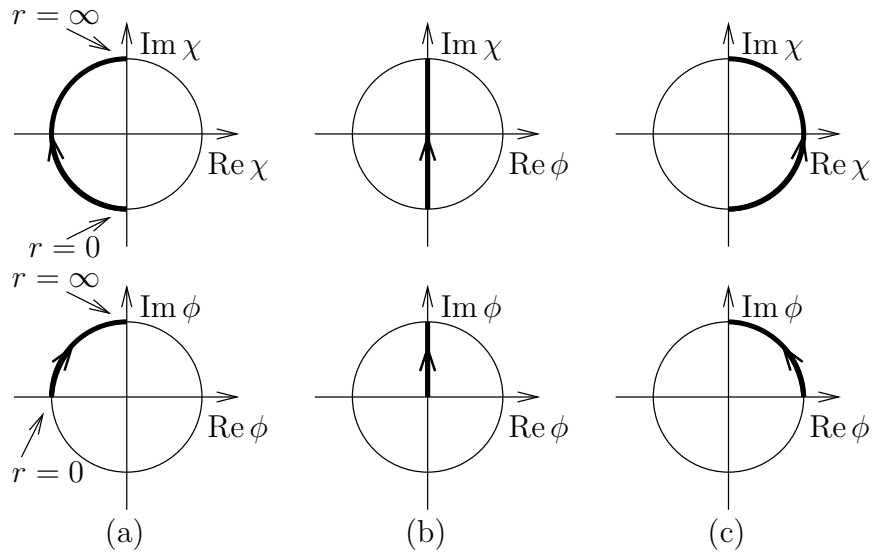}%
    \end{picture}%
    \setlength{\unitlength}{4144sp}%
    \begin{picture}(4974,3261)(664,-2941)
      \put(3151,-2941){\makebox(0,0)[b]{\smash{(b)}}}
      \put(4951,-2941){\makebox(0,0)[b]{\smash{(c)}}}
      \put(1351,-2941){\makebox(0,0)[b]{\smash{(a)}}}
      \put(1441,-1591){\makebox(0,0)[lb]{\smash{$\Imag\phi$}}}
      \put(1846,-2311){\makebox(0,0)[lb]{\smash{$\Real\phi$}}}
      \put(5041,-1591){\makebox(0,0)[lb]{\smash{$\Imag\phi$}}}
      \put(3646,-2311){\makebox(0,0)[lb]{\smash{$\Real\phi$}}}
      \put(5446,-2311){\makebox(0,0)[lb]{\smash{$\Real\phi$}}}
      \put(3241,-1591){\makebox(0,0)[lb]{\smash{$\Imag\phi$}}}
      \put(676,-1411){\makebox(0,0)[lb]{\smash{$r=\infty$}}}
      \put(676,-2581){\makebox(0,0)[lb]{\smash{$r=0$}}}
      \put(3646,-736){\makebox(0,0)[lb]{\smash{$\Real\phi$}}}
      \put(1441,-16){\makebox(0,0)[lb]{\smash{$\Imag\chi$}}}
      \put(1846,-736){\makebox(0,0)[lb]{\smash{$\Real\chi$}}}
      \put(3241,-16){\makebox(0,0)[lb]{\smash{$\Imag\chi$}}}
      \put(5041,-16){\makebox(0,0)[lb]{\smash{$\Imag\chi$}}}
      \put(5446,-736){\makebox(0,0)[lb]{\smash{$\Real\chi$}}}
      \put(676,164){\makebox(0,0)[lb]{\smash{$r=\infty$}}}
      \put(991,-1231){\makebox(0,0)[b]{\smash{$r=0$}}}
    \end{picture}%
  }
  \caption{Topological transition in the $SU(2)$ Higgs model: behavior
    of the fields $\phi$ and $\chi$.  Bold arrows show the change of
    the field as the radial coordinate increases from $r=0$ to
    $r=\infty$.  The configurations are shown: (a) at initial time,
    (b) in the middle of the process and (c) at final time.}
  \label{fig:transition}
\end{figure}

In this gauge no $r$-independent vacuum exists, but transition from
vacua with $n=0$ and $n=1$ is described in a very symmetric way.  The
behavior of the fields $\chi$ and $\phi$ for such transition is shown
in fig.~\ref{fig:transition}.  In the original 4-dimensional model
this topology changing process corresponds to a transition where the
fields wind over the lower hemisphere of $SU(2)$ before the transition
and over the upper hemisphere after the transition.

\paragraph{Boundary conditions.}
The formulation of the $\Theta$ boundary conditions \eqref{BVP_bc2}
requires additional effort in this model, as compared to the case of
single scalar field.  The reason is that one of the fields is not
really a physical field in $A_0=0$ gauge---if fixed at some time, it
can be expressed in terms of other fields at any other moment of time
using Gauss' law, which is a first order equation (in time).  The
corresponding expression is quite complicated, so it is comfortable
instead to solve the second order equations only and impose Gauss' law
at one moment of time.  Then the Gauss' law is automatically satisfied
at all other times.

So, we have to impose $\Theta$ boundary conditions \eqref{BVP_bc2}
only on four of the five fields $a_1$, $\alpha$, $\beta$, $\mu$, $\nu$
(or, to be more precise, on four combinations of these fields).  At
one moment of time we also have to impose Gauss' law constraint and
fix the gauge freedom to make the boundary value problem properly
defined.  The two latter constraints are equations on complex
functions of spatial coordinate $r$ (recall that all fields in our
problem are complex and gauge function $\Omega$ may be complex too).
Half of these equations are not needed---they are duplicated by the
reality conditions at the CD part of the contour.  Reality implies
that imaginary part of Gauss' law is zero and forbids gauge
transformations with imaginary $\Omega$.  So we are left only with the
real part of Gauss' law and have to fix invariance under real gauge
transformations at the initial time moment.  The latter can be done by
setting a certain combination of fields corresponding to unphysical
perturbation of the initial vacuum to zero.
This gives us the needed number of boundary conditions to determine
unique (in general) solution.

\paragraph{Zero mode.} 
One more complication is that, in continuous formulation, the boundary
value problem (\ref{final_BVP}) \emph{does have} an invariance under
translations along real time (both field equations and the boundary
conditions are invariant under such translation).  To properly define
the boundary value problem one has to fix the position of the
solution in time.  In the lattice version this invariance is violated
by the discretization and finite volume effects.

A possible modification of the equations is the following.  One of
the equations \eqref{BVP_bc2} (in lattice case $\bk$ takes discrete
values) is changed to
\[
  |\fk| = \e^{-\Theta} |g_{\bk}| \;.
\]
If the field is in linear regime at initial time relative phase between
$\fk$ and $g_\bk$ is zero because total energy is real.  So in linear
regime we get the solution to the original boundary value problem.

Instead of the equation for the relative phase any (real) equation
which is not invariant under time translations can be used.  It fixes
the position of the solution.  We fix the center of spatial
distribution of field $\chi$ at part A of the contour, which
corresponds to fixing of the position of incoming wave.

\section{Numerical results}\label{sec:num_results}

There are several peculiar features of the boundary value problem
\eqref{final_BVP} that make solving it numerically a computational
challenge.  First, equation \eqref{BVP_eq} is non-linear.  Second, at
$\Theta\ne0$ the fields are necessarily complex and the solution is
not a maximum of \eqref{Fexp} but only a saddle point.  Third, the
time contour contains both Minkowskian (AB and CD) and Euclidean (BC)
parts, so the problem is both of hyperbolic and elliptic type.
Finally, the initial boundary conditions \eqref{BVP_bc2} should be
imposed when the system is already close to linearity which requires
large spatial volume of the configuration.

These factors constrain lattice parameters and applicable numerical
techniques.  In the lattice version the boundary value problem
\eqref{final_BVP} becomes a set of non-linear algebraic equations for
the field values $\varphi_{iJ}=\{a_1,\alpha,\beta,\mu,\nu\}(t_i,r_j)$
at the lattice sites with coordinates $(t_i,r_j)$, where
$r_0,\dots,r_{N}$ corresponds to spatial radial direction,
$t_{-1},\dots,t_{N_t+1}$ are complex time coordinates on contour ABCD,
index $J$ is a combination of the spatial index $j$ and field type
$\{a_1,\alpha,\beta,\mu,\nu\}$ and runs from $1$ to $5N-4$.  The
lattice size is characterized by the lengths of AB, BC and CD parts of
the time contour ($T_{M{\text{initial}}}$, $T_{E}$ and
$T_{M\text{final}}$ respectively), and spatial size $L$.  While on the
Euclidean part the solution is compact in space (and has
characteristic size of order of 1), it is generally evolving along the
light cone in the Minkowskian regions.  This requires that $L\gtrsim
T_{M\text{inital}}$.  In calculations we chose $L=8$ and
$T_{M\text{inital}}=6$ (in units of Higgs boson mass).

The numerical method to solve the set of equations which constitute
the lattice version of the boundary value problem (\ref{final_BVP}) is
to be chosen according to the specifics of the problem described
above.  To get rid of the non-linearity we employ a multidimensional
Newton--Raphson method which
approaches the desired solution iteratively. At each iteration, the
\emph{linearized} equations in the background of the current
approximation are solved. The next approximation is obtained by adding
the solution to the background, and the procedure is repeated. The
advantage of the algorithm is that it does not require
positive-definiteness of the matrix of second derivatives. It is,
however, sensitive to zero modes. In the absence of zero modes, the
algorithm converges quadratically; the accuracy of $10^{-9}$ is
typically reached in 3-5 iterations. The convergence slows down in the
presence of very soft modes, as typically happens near bifurcation
points.

At each Newton-Raphson iteration one solves the set of
$N_t\times 5N$ linear equations of the general form 
\[
  L\cdot u = d,
\]
where $u$ is the vector formed of $N_t\cdot5N$ unknowns, $L$ is the
matrix of dimension $N_t\cdot5N\times N_t\cdot5N$ (first variation of
the full non-linear equations) and $d$ is a constant vector (full
equations evaluated at the current background; at the desired solution
$d=0$).  The inversion of this matrix is the most time consuming part
of the calculation; its efficiency determines how large $N_t$ and $N$
can be used.  The matrix $L$ is neither positive-definite nor even
symmetric, but has a special sparse structure as it originates from
the second order differential equation.  The linear equations relate
only adjacent time slices $i-1$, $i$, $i+1$, so one can eliminate
equations for alternate time slices
\cite{Bonini:1999kj,Bonini:1999fc}.  Moreover, one can do that in
parallel, making use of the power of multiprocessor computers.  The
algorithm requires $\sim N_t(5N)^3$ multiplications.  Note, that it is
highly asymmetric in $N_t$ and $N$---one can use large $N_t$ but is
very constrained with the choice of $N$.  The calculations were
performed with $N=64$ and $N_t=350$.

The Newton-Raphson method requires a good initial approximation for
the solution.  This favors the following general strategy. We first
find the periodic instanton solution (which corresponds to $\Theta=0$)
\cite{Bonini:1999fc}.  After the periodic instanton is found, we
change parameters $T$ and $\Theta$ by small steps, using the solution
from the previous run as a starting configuration for the next one.
At each step we calculate the energy $E$, number of particles $N$ and
the exponential suppression factor $F(\varepsilon,\nu)$.

\begin{figure}[tb]
  \centerline{\psfig{width=290pt,file=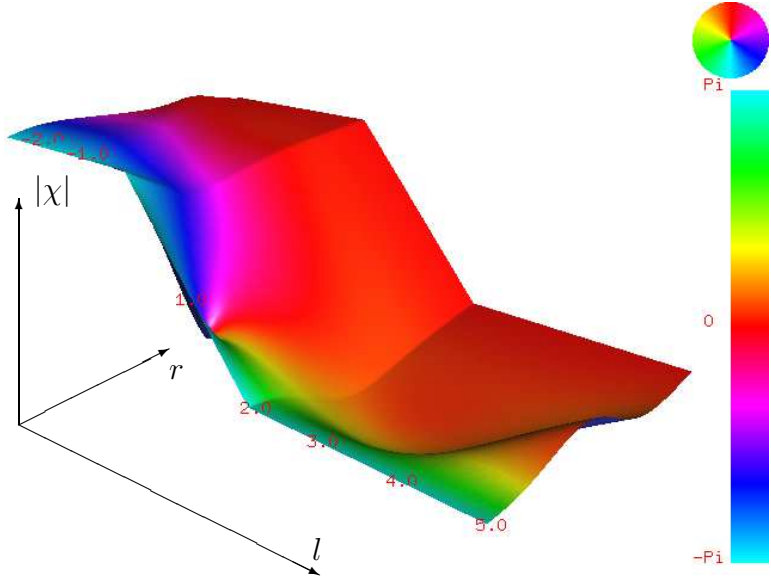}%
    \begin{picture}(0,0)
      \unitlength=1cm
      \put(-10,2){\vector(0,1){3}}\put(-9.8,5){$|\chi|$}
      \put(-10,2){\vector(2,1){2}}\put(-6.1,0.2){$l$}
      \put(-10,2){\vector(2,-1){4}}\put(-8,2.6){$r$}
    \end{picture}%
  }
  \caption{Configuration for $T/2=1.98$ and $\Theta=2.74$.
    $\varepsilon=0.711$.
    Height and color represent $|\chi|$ and the phase of $\chi$,
    respectively.  For visualization purposes the Euclidean part of
    the contour is inclined.}
  \label{fig:conf_good}
\end{figure}

A typical configuration is shown in figure~\ref{fig:conf_good}.  One
can clearly see that the phase of the field $\chi$ behaves as shown in
figure~\ref{fig:transition}, going from $-\pi$ at $r=0$ to $\pi$ at
$r=L$ along different sides of the circle in the initial and final
states.  In the middle of the Euclidean region, a zero of the field
$\chi$ is present (center of the ``instanton'').  Also the
incoming/outgoing wave is observed at initial/final Minkowskian part.

It is fairly straightforward to check that the singularities shown in
figure~\eqref{fig:time_contour} are indeed present, by continuing the
solution from the part BC of the contour to the whole complex time
plane.  The left singularity is approximately at the same distance
from the Euclidean part of the contour for all solutions.  The right
singularity moves to larger positive times as the energy increases.

\begin{figure}[tb]
  \psfig{file=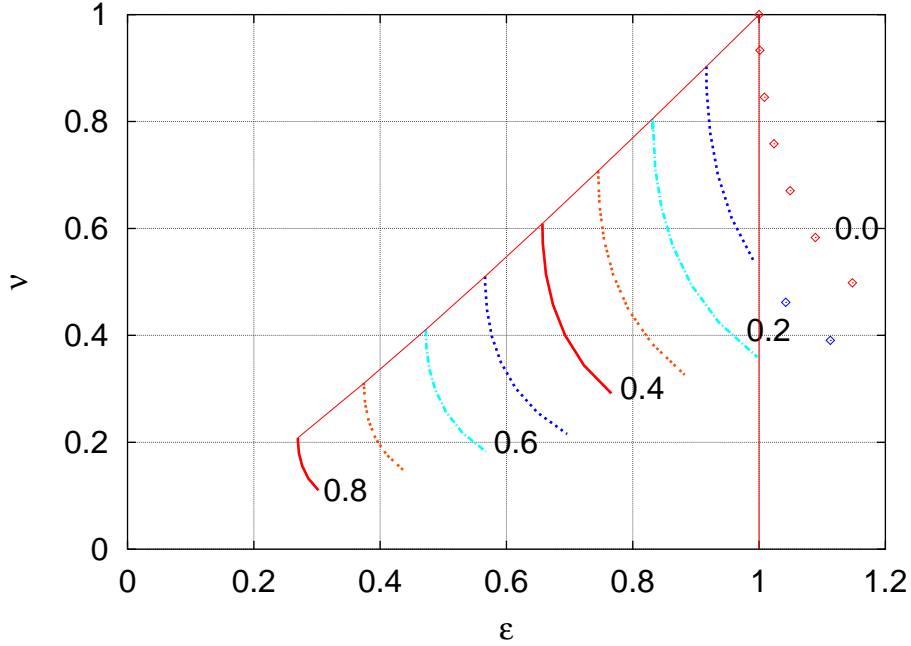}
  \caption{Lines of $F(\varepsilon,\nu)=\const$.  $F$ is normalized to
    be unity at $\nu=\varepsilon=0$ (instanton).  Diagonal line
    directed from sphaleron ($\nu=\varepsilon=1$) towards the origin
    is the line of periodic instantons.  Open symbols correspond
    to configurations with wrong topology.  Numbers near
    the curves show the suppression factor $F$.}
  \label{fig:ne}
\end{figure}

Summary of the results is given in figure~\ref{fig:ne}.  Lines of
constant suppression factor $F$ are shown on $\varepsilon$--$\nu$
plane.  Units are normalized to the sphaleron:
$\nu_\mathrm{sph}=\varepsilon_\mathrm{sph}=1$.

Behavior of the constant suppression lines has the following features.
Near the periodic instanton ($\Theta=0$), the dependence of $F$ on
$\nu$ is weak.  This can be seen analytically from the boundary
problem \eqref{final_BVP} itself.  Making use of the fact that $F$ is
stationary with respect to $T$ and $\Theta$ one finds
\[
  \left.\frac{dN}{dE}\right|_{F=\const} = -\frac{T}{\Theta} \;,
\]
which is infinite as $\Theta\to0$.

When one moves away from the periodic instantons, lines of constant
$F$ flatten out; in other words, increase of energy in this region
leads to smaller decrease of the suppression exponent than in the
vicinity of the periodic instanton.

\begin{figure}[tb]
  \psfig{file=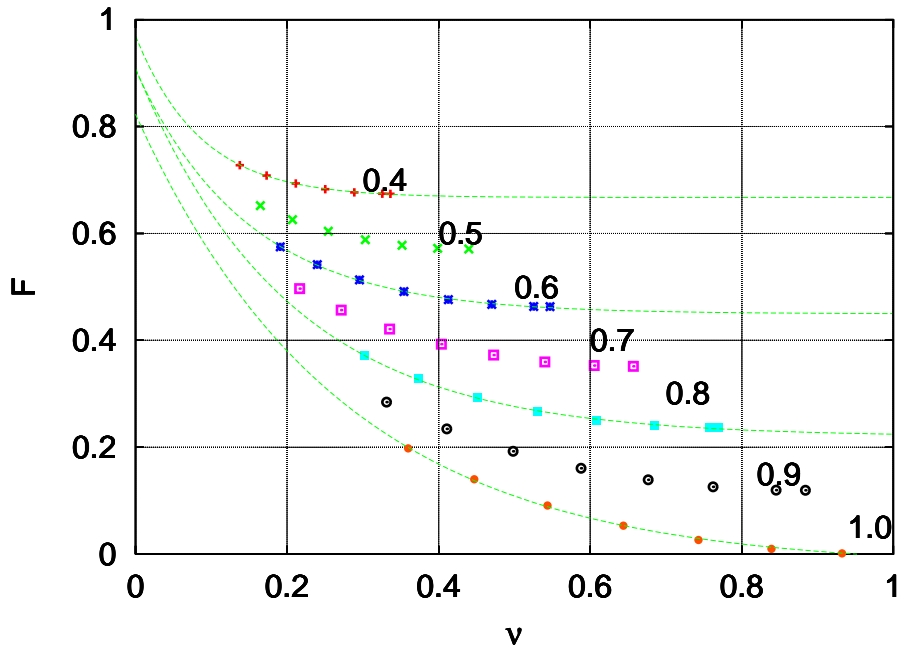}
  \caption{Dependence of the suppression factor $F$ on $\nu$ for
    different energies.  Numbers near the curves show values of
    $\varepsilon$.  Lines are extrapolations of data with functions of
    the form $a e^{-b\nu}-c$.}
  \label{fig:fn}
\end{figure}

The quantity of primary interest is the two-particle cross section
\[
  \sigma_\mathrm{tot}(E) \sim
    \exp\left\{
      -\frac{16\pi^2}{g^2}F\left({E}/{E_\mathrm{sph}},0\right)
    \right\}
\]
We can plot $F(\varepsilon,\nu)$ as a function of $\nu$ for different
values of energy (fig.~\ref{fig:fn}).  Extrapolation of the data for
$\varepsilon=0.4,\dots,1.0$ crosses the line $\nu=0$ at
$F\sim0.8,\dots,1$.  So we conclude that for energies below the
sphaleron, the exponent of the suppression factor of the tunneling
processes is only about 20\% less than for zero energy (instanton
case).

Finally, the open symbols in fig.~\ref{fig:ne} correspond to
evolutions where the field configuration, after having undergone a
topological transition (marked by the zero of the $\chi$ field on BC
part of the time contour---see
fig.~\ref{fig:conf_good}), goes trough a reverse transition on the CD
part of the contour and
returns to the original topological sector.  We will discuss these
solutions in the next section.

\section{Conclusions and Outlook}\label{sec:conclusions}

Of course, the point of major interest is whether at sufficiently high
energy particle collisions can induce unsuppressed topological
transitions and baryon number violation in the Electroweak Theory.  In
terms of the graph of fig.~\ref{fig:ne} the question is whether the
line corresponding to $F(\varepsilon,\nu)=0$ does approach the $\nu=0$
axis asymptotically for increasing $\varepsilon$ and, if it does, at 
what rate.

It should be noted that in our study we have not yet been able to
obtain topology changing solutions with $F=0$.  As we described, the
exploration of the space of solutions is done through a deformation
procedure, by which the parameters $T$ and $\Theta$ are gradually
changed.  This in turn moves the solutions in the $\varepsilon,\nu$
plane.  We observed that as one goes beyond the sphaleron energy, the
solutions become unstable.  The field configurations linger for longer
and longer times in the neighborhood of the sphaleron and then return
to the original topological sector.  One thus obtains solutions
satisfying the equations of motion and all the boundary conditions,
but for the requirement that the topology changes.  This instability
is not unexpected.  It was observed by the authors of
ref.~\cite{Bonini:1999kj} in their study of transitions across a
potential barrier in a quantum mechanical model.  It indicates a
likely bifurcation in the space of semiclassical solutions as one
reaches the barrier energy.  In ref.~\cite{Bonini:1999kj} the problem
was solved by deforming the time contour for the evolution (see
fig.~\ref{fig:time_contour}) to make it go below the real axis (the
segment CD in fig.~\ref{fig:time_contour}) before returning to the
real time axis.  This helped pinpoint a set of solutions with
transition across the barrier.  We are pursuing a similar strategy in
our current search.  The computational problems one faces are
nevertheless daunting.  The most crucial factor is the ability of
following the solutions for negative and positive time well into the
linear regime, where they are settled in the two different different
topological sectors.  This in turn requires the use of a grid of large
extent in the radial direction and makes the computation quite
demanding.  We are making progress and hope that we will be able to
report on topology changing solutions above the sphaleron energy in
the near future.  For the moment, we believe that the detailed
information we obtained for the lines of constant suppression factor
below the sphaleron energy can also be of value.  In particular, the
marked bending of the lines toward increasing energy as one lowers the
incoming particle number seems to indicate that topology changing
transitions in particle collisions will occur, if at all, only for
energy substantially higher than $E_\mathrm{sph}$.

\section{Acknowledgements}\label{sec:acknowledgements}

The authors are grateful to A.Kuznetsov for numerous discussions at
different stages of the work.

This work has supported in part under DOE grant DE-FG02-91ER40676 
and by Award RP1-2103 of the U.S. Civilian
Research~\& Development Foundation (CRDF).  F.B.~and V.R.~are
supported also by Russian Foundation for Basic Research (RFBR) grant
99-01-18410; Council for Presidential Grants and State Support of
Leading Scientific Schools, grant 00-15-96626.  P.T.~is supported by
the Swiss Science Foundation, grant 21-58947.99.

\pagebreak[4]

\end{document}